\documentclass[12pt]{iopart}
\usepackage{iopams}
\usepackage{graphicx}
\begin{document}

\title[The influence of magnetic sublattice dilution on magnetic order in CeNiGe$_3$ and UNiSi$_2$]{The
influence of magnetic sublattice dilution on magnetic order in CeNiGe$_3$ and UNiSi$_2$}

\author{A P Pikul}

\address{Institute of Low Temperature and Structure Research, Polish Academy of Sciences,
P~Nr~1410, 50--590~Wroc{\l}aw~2, Poland}
\ead{A.Pikul@int.pan.wroc.pl}
\begin{abstract}
Polycrystalline samples of the Y-diluted antiferromagnet CeNiGe$_3$ ($T_{\rm N}$~=~5.5~K) and
Th-diluted ferromagnet UNiSi$_2$ ($T_{\rm C}$~=~95~K) were studied by means of x-ray powder
diffraction, magnetization and specific heat measurements performed in a wide temperature range.
The lattice parameters of the Ce$_{1-x}$Y$_x$NiGe$_3$ alloys decrease linearly with increasing the
Y content, while the unit cell volume of U$_{1-x}$Th$_x$NiSi$_2$ increases linearly with rising the
Th content. The ordering temperatures of the systems decrease monotonically with increasing $x$
down to about 1.2~K in Ce$_{0.4}$Y$_{0.6}$NiGe$_{3}$ and 26~K in U$_{0.3}$Th$_{0.7}$NiSi$_2$,
forming a dome of a long-range magnetic order on their magnetic phase diagrams. The suppression of
the magnetic order is associated with distinct broadening of the anomalies at $T_{\rm N, C}$ due to
crystallographic disorder being a consequence of the alloying. Below the magnetic percolation
threshold $x_{\rm c}$ of about 0.68 and 0.75 in the Ce- and U-based alloys, respectively, the
long-range magnetic order smoothly evolves into a short-range one, forming a tail on the magnetic
phase diagrams. The observed behaviour Ce$_{1-x}$Y$_x$NiGe$_3$ and U$_{1-x}$Th$_x$NiSi$_2$ is
characteristic of diluted magnetic alloys.
\end{abstract}

\submitto{\JPCM}

\section{Introduction}

Partial substitution of some chemical elements by another ones, is one of the methods of modifying
and examining the ground states of intermetallic compounds with unique physical properties. Such
alloying (or doping) is particularly fruitful in strongly correlated electron systems, in which
different sizes and electron configurations of the swapped atoms may significantly influence the
exchange integral, being one of the crucial parameters determining magnetic properties of these
systems \cite{stewart1, stewart2, loehneysen}. Additionally, the chemical doping introduce some
structural disorder, which can be another source of non-trivial magnetic behaviour \cite{miranda}.

In the course of our studies of the influence of the chemical substitution on physical properties
of selected dense Kondo systems \cite{pikul7, westerkamp, pikul8, pikul9, pikul10, pikul11}, we
have performed alloying studies of two ternary intermetallic phases, namely CeNiGe$_3$ and
UNiSi$_2$, crystallizing in two orthorhombic and -- to some extent -- similar structures of the
SmNiGe$_3$- \cite{salamakha} and CeNiSi$_2$-type \cite{akselrud}, respectively. CeNiGe$_3$ orders
antiferromagnetically below the N{\'e}el temperature $T_{\rm N}$~=~5.5~K and exhibits features of a
dense Kondo system with the Kondo temperature $T_{\rm K}$ being close to $T_{\rm N}$ \cite{pikul1}.
Physical properties measurements performed under hydrostatic pressure revealed that the N\'{e}el
temperature of CeNiGe$_3$ initially increases with increasing pressure up to about 8~K at $P_{\rm
max}\approx$~3~GPa. At higher pressure $T_{\rm N}$ rapidly decreases and the antiferromagnetic
order is suppressed at the critical pressure $P_{\rm c}$~=~5.5~GPa. Simultaneously, the
$4f$-electrons of cerium become partly delocalized and at about 8~GPa the compound exhibits
features of an intermediate valence system. Most importantly, at the critical pressure CeNiGe$_3$
becomes superconducting below $T_{\rm c}$~=~0.48~K \cite{nakashima1,kotegawa1} and -- as revealed
by nuclear quadrupole resonance measurements -- the superconductivity has an unconventional
character resulting from the presence of the $4f$ electrons of cerium
\cite{harada1,harada2,harada3}. Interestingly, also in YNiGe$_3$, being an isostructural
non-$f$-electron reference to CeNiGe$_3$, a superconducting phase transition was found below about
0.48~K, yet at ambient pressure \cite{pikul6}. UNiSi$_2$ exhibits in turn a ferromagnetic order of
nearly localized magnetic moments of uranium ($T_{\rm C}$~=~95~K) and some features characteristic
of Kondo lattices \cite{kaczorowski2,pikul11}. Our investigations of the solid solutions
UNi$_{1-x}$Co$_x$Si$_2$ ($0 \leqslant x \leqslant 1$) indicated a very robust nature of the
ferromagnetism observed in the parent compound UNiSi$_2$ -- although the ferromagnetic ordering is
gradually suppressed upon stepwise substitution of Ni by Co, the phase transition exhibits clearly
the ferromagnetic character up to $x=0.96$ \cite{pikul12}. Recent pressure experiments performed on
UNiSi$_2$ showed that the ferromagnetism of the compound is quenched above 5.5~GPa, where the
$5f$-electrons of uranium become delocalized \cite{sidorov}.

In order to shed more light on the intriguing properties of CeNiGe$_3$ and UNiSi$_2$ we have
investigated physical properties of their solid solutions with their non-magnetic isostructural
counterparts YNiGe$_3$ and ThNiSi$_2$, respectively. Please note, that LaNiGe$_3$, which would be a
more appropriate reference for CeNiGe$_3$, most probably does not exist \cite{pikul1}. In this
paper we present results of the magnetic susceptibility and specific heat measurements, indicating
suppression of the long-range magnetic order and its smooth evolution into a short-range one.

\section{Experimental details}

Polycrystalline samples of the alloys Ce$_{1-x}$Y$_x$NiGe$_3$ and U$_{1-x}$Th$_x$NiSi$_2$ ($0
\leqslant x \leqslant 1$) were prepared by conventional arc melting the stoichiometric amounts of
the elemental components in protective atmosphere of an argon glove box. The pellets with Ce were
subsequently wrapped in a molybdenum foil, sealed in an evacuated silica tube, and annealed at
800$^{\circ}$C for one week, while the U-based samples were wrapped in a tantalum foil annealed for
four weeks. The quality of the products was checked by means of x-ray powder diffraction, which
showed that all the samples were single phases.

Magnetic properties of the alloys were studied using a commercial Quantum Design Magnetic Property
Measurement System at temperatures 1.7--400~K and in magnetic fields up to 5~T. The heat capacity
was measured at temperatures ranging from 360~mK up to room temperature, using a Quantum Design
Physical Property Measurement System.

\section{Results}

\subsection{Crystal structure}

\begin{figure}
\centering
\includegraphics[width=7cm]{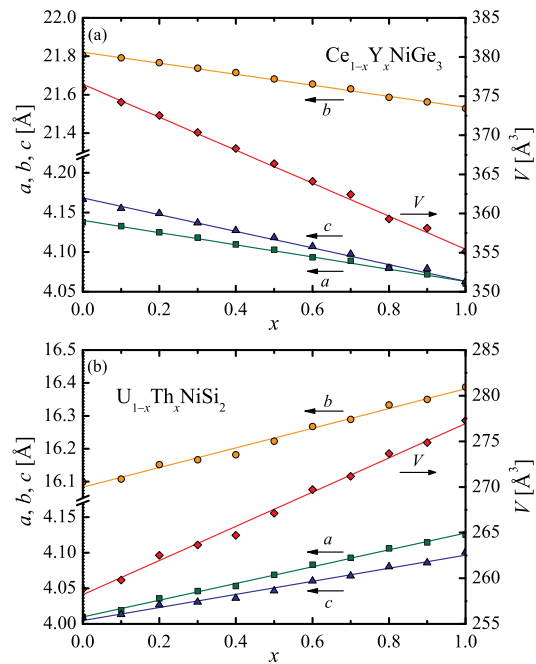}
\caption{\label{lattice_parameters} Lattice parameters $a$, $b$ and $c$ and unit cell volume $V$
of the Ce$_{1-x}$Y$_x$NiGe$_3$ (a) and U$_{1-x}$Th$_x$NiSi$_2$ (b) alloys as a function of the nominal
yttrium and thorium contents $x$, respectively. Solid lines represent linear fits to the experimental data.}
\end{figure}

Analysis of the x-ray powder diffraction patterns (not presented here) confirmed that CeNiGe$_3$
and YNiGe$_3$ crystallize in the SmNiGe$_3$-type structure (space group \emph{Cmmm}
\cite{salamakha}), while UNiSi$_2$ and ThNiSi$_2$ -- in the CeNiGe$_2$-type structure (space group
\emph{Cmcm} \cite{akselrud}). The refined lattice parameters are $a$~=~4.1377(5)~\AA,
$b$~=~21.807(3)~\AA~ and $c$~=~4.1668(5)~\AA~ for CeNiGe$_3$, $a$~=~4.0604(5)~\AA,
$b$~=~21.529(2)~\AA~ and $c$~=~4.0627(3)~\AA~ for YNiGe$_3$, $a$~=~4.0097(5)~\AA,
$b$~=~16.099(2)~\AA, and $c$~=~4.0089(5)~\AA~ for UNiSi$_2$, and $a$~=~4.1256(4)~\AA,
$b$~=~16.388(2)~\AA, and $c$~=~4.1009(5)\AA~ for ThNiSi$_2$, all being close to the values reported
in the literature \cite{salamakha,akselrud,pikul2,kaczorowski2}. The Rietveld refinement revealed
that partial substitution of Ce by smaller Y and U by larger Th does not change the crystal
structures of the systems but results -- respectively -- in linear compression and expansion of the
lattice. As can be inferred from figure~\ref{lattice_parameters}, the observed change of the
unit-cell volume (in total up to -6\% in the Ce phases and +7\% in the U alloys) is in both systems
monotonic and linear in $x$ in each principal crystallographic direction.

\subsection{Magnetic properties}

\begin{figure}
\centering
\includegraphics[width=7cm]{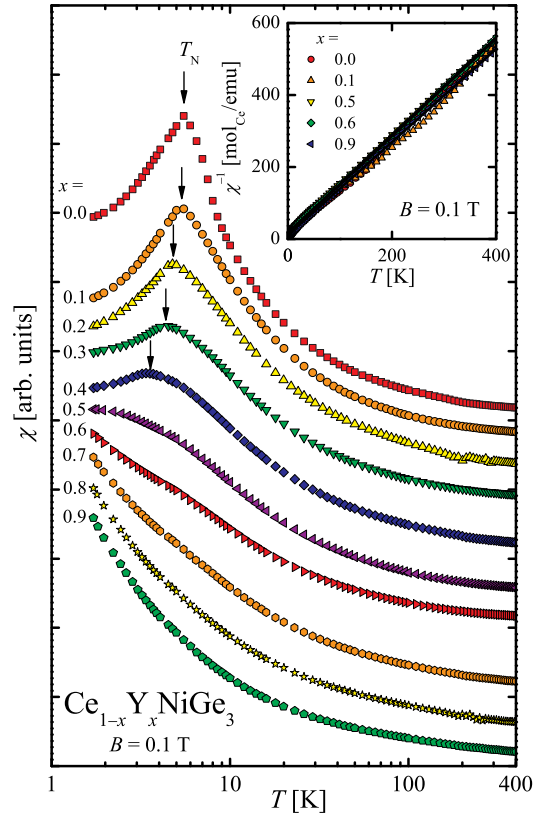}
\caption{\label{magnetic_properties_CYNG} Temperature variations of the magnetic susceptibility $\chi$
of Ce$_{1-x}$Y$_x$NiGe$_3$ in a logarithmic scale.
The curves are shifted upward for the sake of clarity; the arrows mark the antiferromagnetic phase
transition temperatures $T_{\rm N}$. Inset: $\chi^{-1}$ vs. $T$ for selected
compositions. Straight solid lines represent the Curie-Weiss fits.}
\end{figure}

The magnetic susceptibility measurements performed for the parent compounds CeNiGe$_3$ and
YNiGe$_3$ confirmed the previous findings \cite{pikul1,nakashima1,tateiwa}. In particular,
CeNiGe$_3$ exhibits localised antiferromagnetism below 5.5~K, while YNiGe$_3$ is an itinerant
paramagnet with nearly temperature independent magnetic susceptibility $\chi(T)$ (not shown here).
Figure~\ref{magnetic_properties_CYNG} displays $\chi (T)$ of Ce$_{1-x}$Y$_x$NiGe$_3$. As is
apparent from the inset to Fig.~\ref{magnetic_properties_CYNG}, the $\chi^{-1}(T)$ curves exhibit
linear behavior above about 100~K and -- as normalised per mole of Ce atoms -- are nearly
superimposed onto each other. Least square fits of the Curie--Weiss law (i.e. $\chi(T) = (1/8)
\mu_{\rm eff}^2/(T-\theta_{\rm p})$) to the experimental data (see the solid lines in the inset to
Fig.~\ref{magnetic_properties_CYNG}) yielded for each alloy the effective magnetic moment $\mu_{\rm
eff}$ of about 2.5(1)~$\mu_{\rm B}$ and the paramagnetic Curie temperature $\theta_{\rm p}$ ranging
between $-20(5)$ and $-10(5)$~K. The values of $\mu_{\rm eff}$ are close to the theoretical
Hund's-rules magnetic moment calculated for a free Ce$^{3+}$ ion (i.e. 2.54~$\mu_{\rm B}$),
pointing to the presence of well localized cerium magnetic moments in all the alloys studied. The
negative $\theta_{\rm p}$ values are in line with antiferromagnetic exchange interactions, and
their relatively large absolute values (in comparison to $T_{\rm N}$) suggest the presence of the
Kondo interactions in the alloys. Due to significant scatter in the $\theta_{\rm p}$ values (not
shown here), no clear evolution was found in the $\theta_{\rm p} (x)$ dependence. The deviation of
$\chi^{-1}(T)$ from the linear behavior below 100~K can be ascribed to thermal depopulation of the
crystalline electric field levels of the cerium ions. Upon increasing $x$, the cusp visible in
$\chi (T)$ of Ce$_{1-x}$Y$_x$NiGe$_3$ at $T_{\rm N}$ moves towards lower temperatures and
significantly broadens. For $x$~=~0.50 the maximum is hardly visible in the magnetic susceptibility
above 2~K (Fig.~\ref{magnetic_properties_CYNG}).

\begin{figure}
\centering
\includegraphics[width=7cm]{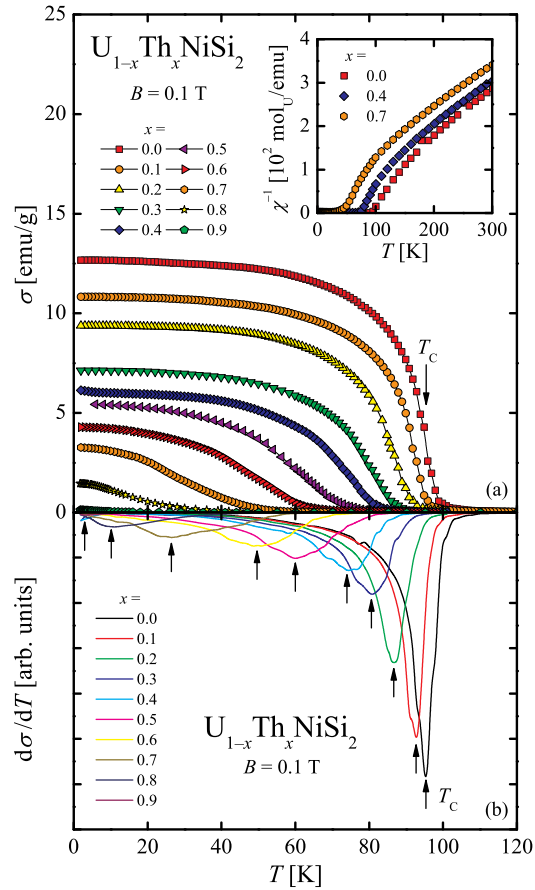}
\caption{\label{magnetic_properties_UTNS} (a) Magnetization $\sigma$ of U$_{1-x}$Th$_x$NiSi$_2$
as a function of temperature $T$ measured in constant magnetic field $B$ in a field-cooling regime;
the arrow marks the ferromagnetic ordering temperature $T_{\rm C}$. Inset: $\chi^{-1}$ vs. $T$ for
selected compositions. (b) Temperature derivative of $\sigma (T)$;
the arrows mark the ordering temperatures.}
\end{figure}

Figure \ref{magnetic_properties_UTNS} presents the magnetic properties of the
U$_{1-x}$Th$_x$NiSi$_2$ alloys. UNiSi$_2$ orders ferromagnetically at the Curie temperature $T_{\rm
C}$~=~95~K (as reported earlier~\cite{kaczorowski2}), while its isostructural non-$f$-electron
reference compound ThNiSi$_2$ is a weak Pauli-like paramagnet. At elevated temperatures, the
$\chi^{-1} (T)$ curves appeared to have similar shape and to be nearly parallel to each other (see
the inset to Fig.~\ref{magnetic_properties_UTNS}). Such a behaviour suggests that the effective
magnetic moment of uranium does not change significantly upon increasing $x$, yet its exact value
can not be estimated from the Curie--Weiss fits (due to partial delocalization of the $5f$
electrons and the crystalline electric field splitting exceeding the temperature range studied). As
apparent in figure \ref{magnetic_properties_UTNS}(a), the initial Brillouin-shaped $\sigma (T)$
curve significantly broadens and flattens upon diluting the magnetic sublattice and $T_{\rm C}$
defined as a position of the minimum in temperature derivative of $\sigma (T)$ (figure
\ref{magnetic_properties_UTNS}(b)) decreases monotonically with increasing $x$.

\subsection{Thermodynamic properties}

\begin{figure}
\centering
\includegraphics[width=7cm]{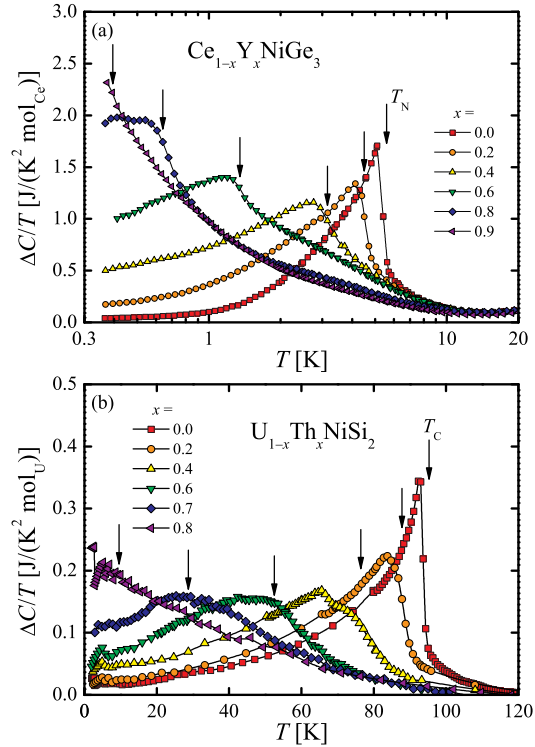}
\caption{\label{thermodynamic_properties_CYNG_UTNS} Temperature dependence of a non-phononic contribution $\Delta C$
to the total specific heat of selected Ce$_{1-x}$Y$_{x}$NiGe$_{3}$ (a) and U$_{1-x}$Th$_x$NiSi$_2$ (b) alloys,
normalized per mole of cerium or uranium, respectively, and divided by $T$. The solid lines serve as guides for the eye
and the arrows mark the phase transition temperatures from temperature derivatives of $\Delta C(T)/T$.}
\end{figure}

Figure \ref{thermodynamic_properties_CYNG_UTNS}(a) displays the low temperature variation of the
$4f$-electron contribution $\Delta C$ to the total specific heat of Ce$_{1-x}$Y$_{x}$NiGe$_{3}$,
estimated by subtracting the phonon specific heat of YNiGe$_3$ (cf. \cite{pikul1}), normalized per
mole of cerium and divided by $T$. As seen, the antiferromagnetic ordering of CeNiGe$_3$ at $T_{\rm
N}$~=~5.5~K manifests itself as a distinct $\lambda$-shaped anomaly in $\Delta C(T)/T$, being in
agreement with the previous reports \cite{pikul1,pikul5}. Upon diluting the Ce-sublattice with Y,
the phase transition anomaly significantly broadens and its position (here defined as a minimum in
the second derivative of $\Delta C(T)/T$; not presented here) moves to 1.35~K for $x$~=~0.6.
Finally, in the most diluted alloy studied, i.e. Ce$_{0.1}$Y$_{0.9}$NiGe$_3$, only a very weak
signature of the phase transition can be found in temperature derivative of $\Delta C(T)/ T$ at
about 0.39~K.

Similar behaviour has been found in the temperature dependence of the $5f$-electron contribution to
the specific heat of U$_{1-x}$Th$_x$NiSi$_2$, obtained by subtracting the phonon contribution of
ThNiSi$_2$ (figure \ref{thermodynamic_properties_CYNG_UTNS}(b)). In particular, the distinct
$\lambda$-shaped anomaly in $\Delta C(T)/T$ manifests the ferromagnetic ordering of the parent
compound at $T_{\rm C}$~=~95~K, being in accord with its magnetic properties \cite{kaczorowski2}.
Partial substitution of U by Th lowers the Curie temperature down to about 29~K in
U$_{0.3}$Th$_{0.7}$NiSi$_2$ and the anomaly at $T_{\rm C}$ quickly evolves into a broad hump. In
the alloy with $x=0.8$ only a weak and blurred anomaly can be found in the specific heat data at
about 11~K.

\begin{figure}
\centering
\includegraphics[width=7cm]{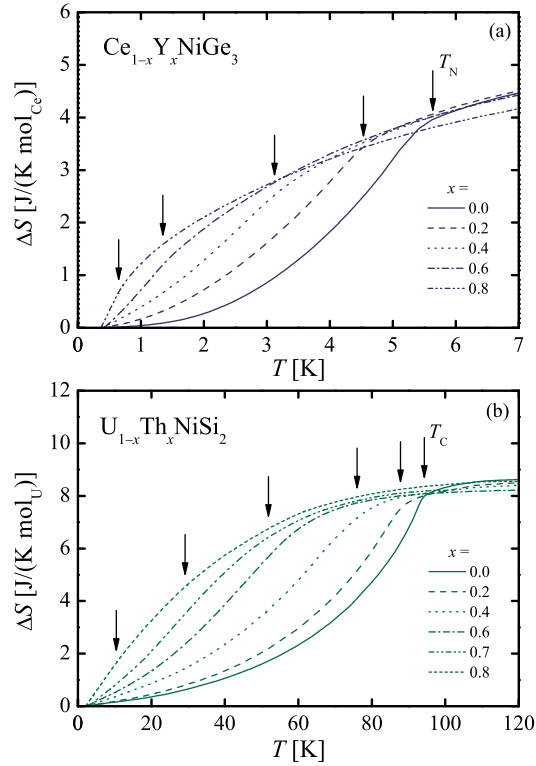}
\caption{\label{entropy} Increase of magnetic entropy $\Delta S$ of selected Ce$_{1-x}$Y$_{x}$NiGe$_{3}$ (a)
and U$_{1-x}$Th$_x$NiSi$_2$ (b) alloys as a function of temperature $T$. The arrows mark the N\'{e}el ($T_{\rm N}$)
and Curie ($T_{\rm C}$) temperatures from $\Delta C(T)/T$ (figure \ref{thermodynamic_properties_CYNG_UTNS}).}
\end{figure}

Figure \ref{entropy} displays the temperature variation of the increase of the magnetic entropy,
defined as $\Delta S (T)= \int_{T_{\rm min}}^T \frac{\Delta C}{T} dT$ (where $T_{\rm min}$ stands
for the low-temperature limit of the specific heat measurements). In CeNiGe$_3$ (figure
\ref{entropy}(a)), $\Delta S$ at $T_{\rm N}$ is smaller than $R\ln{2}$ ($\approx$~5.76~J/(mol K)),
being the expected value for a disordered ground doublet. The reduction of the entropy is most
probably due to the existence of short-range interactions above $T_{\rm N}$ (well visible as a
characteristic tail in $\Delta C(T)/T$ above $T_{\rm N}$; see figure
\ref{thermodynamic_properties_CYNG_UTNS}(a)) and the Kondo effect (evidenced previously
\cite{pikul1}). In UNiSi$_2$ (figure \ref{entropy}(b)), $\Delta S$ at $T_{\rm C}$ is much larger
than $R\ln{2}$, which points out at significant contribution of the excited crystalline electric
field levels to the magnetic entropy at the ordering temperature.

In the paramagnetic region, the $\Delta S(T)$ curves calculated for both the
Ce$_{1-x}$Y$_{x}$NiGe$_{3}$ (figure \ref{entropy}(a)) and U$_{1-x}$Th$_x$NiSi$_2$ alloys (figure
\ref{entropy}(b)) nearly superimpose onto each other, suggesting that the entropy contribution of
the excited crystalline electric field levels is concentration independent in both systems studied.
In other words, the crystalline electric field in Ce$_{1-x}$Y$_{x}$NiGe$_{3}$ and
U$_{1-x}$Th$_x$NiSi$_2$ is only weakly altered by the dilution of the magnetic sublattices, which
is in line with the magnetic properties of the alloys. In the ordered region in turn, the
temperature dependencies of the magnetic entropy of both systems strongly vary with $x$. In
particular, upon increasing the Y/Th content up to $x = 0.2$, the kinks visible in $\Delta S(T)$ of
CeNiGe$_3$ and UNiSi$_2$ at the respective ordering temperatures move towards lower temperatures.
For larger $x$ the sharp anomalies at $T_{\rm N, C}$ evolve into broad, featureless curves.

\section{Discussion}

\begin{figure}
\centering
\includegraphics[width=7cm]{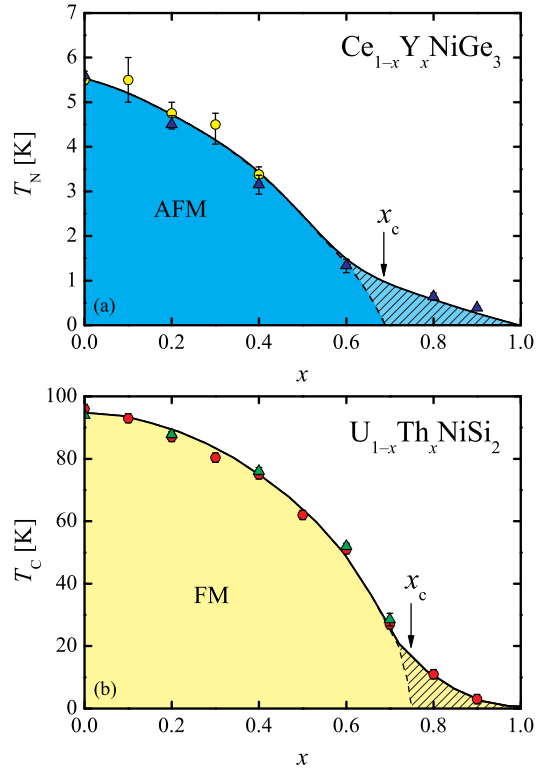}
\caption{\label{phase_diagram} Tentative magnetic phase diagrams for Ce$_{1-x}$Y$_{x}$NiGe$_{3}$ (a)
and U$_{1-x}$Th$_x$NiSi$_2$ (b). Circles correspond to the maxima in $\chi(T)$ or to the minima in
temperature derivative of $\sigma (T)$. Triangles represent the minimum in the temperature derivative
of $\Delta C(T)/T$. Solid lines serve as guides for the eye and the arrows mark the critical concentrations.}
\end{figure}

Figure \ref{phase_diagram} displays tentative magnetic phase diagrams for the
Ce$_{1-x}$Y$_{x}$NiGe$_{3}$ and U$_{1-x}$Th$_x$NiSi$_2$ alloys, constructed on the basis of the
so-far obtained experimental data. The behaviour of both systems reminds those reported for diluted
magnetic alloys \cite{mydosh}. In particular, in the Ce/U-rich part of the diagrams one can
distinguish a dome of a long-range magnetic order, followed by a characteristic tail of a
short-range magnetic order in the diluted limit. These two concentration regimes are separated from
each other by the critical concentration $x_{\rm c}$, roughly estimated by extrapolation of the
initial slope of $T_{\rm N, C}(x)$ down to absolute zero temperature, as being of about 0.64 and
0.75 in Ce$_{1-x}$Y$_{x}$NiGe$_{3}$ and U$_{1-x}$Th$_x$NiSi$_2$, respectively.

The long-range character of the magnetic order of the alloys with $x$~=0.0 and 0.2 can be deduced
\emph{i.a.} from the presence of pronounced anomalies in temperature dependencies of the magnetic
entropy (figure \ref{entropy}). In particular, since the shape of $\Delta S(T)$ is closely related
to the magnetic structure and the thermal demagnetization processes, simple magnetic structures
with high anisotropy yield sharp increase in the entropy with a distinct kink at the ordering
temperature (cf. \cite{marcano} and references therein). Upon rising $x$, an increasing degree of
crystallographic disorder (introduced by the alloying) results in smearing the anomaly at $T_{\rm
N, C}$ (figure \ref{entropy}). Finally, for large values of $x$, any kink in $\Delta S (T)$ is
hardly visible, although some weak anomalies are still visible in other physical characteristics
studied (cf. figures \ref{magnetic_properties_CYNG}, \ref{magnetic_properties_UTNS},
\ref{thermodynamic_properties_CYNG_UTNS}). The latter fact, together with the change of the
curvature in $T_{\rm N, C}(x)$ at $x_{\rm c}$ (figure \ref{phase_diagram}), suggests a domination
of short-range interactions at $x > x_{\rm c}$ (cf. \cite{mydosh}).

The observed evolution of the magnetic properties of the Ce$_{1-x}$Y$_{x}$NiGe$_{3}$ and
U$_{1-x}$Th$_x$NiSi$_2$ alloys can be roughly explained as resulting from a nature of
indirect-exchange Ruderman-Kittel-Kasuya-Yosida interactions, which are responsible for the onset
of a magnetic order in metals. Since these interactions are a function of a distance between the
localized magnetic moments, the random occupation of the Ce/Y and U/Th sites by the $f$-electrons
leads most probably to an appearance in the most diluted alloys of magnetic clusters of different
sizes and ordering temperatures \cite{mydosh}. As a consequence, the observed anomalies in the
physical properties studied become more and more blurred with increasing $x$, and $T_{\rm N, C}$
represent a mean ordering temperature of the particular alloys. Since the size of the clusters
remains finite far below the percolation threshold, a characteristic tail in the magnetic phase
diagrams is visible (figure \ref{phase_diagram}). The latter hypothesis need to be verified by
further experiments, e.g. AC magnetic susceptibility, neutron diffraction etc.

\section{Summary}

The collected experimental data revealed that the dilution of the magnetic sublattices of
CeNiGe$_3$ ($T_{\rm N}$~=~5.5~K) and UNiSi$_2$ ($T_{\rm C}$~=~95~K) leads to rapid decrease of
their ordering temperatures down to about 1.35~K in Ce$_{0.4}$Y$_{0.6}$NiGe$_{3}$ and 29~K in
U$_{0.3}$Th$_{0.7}$NiSi$_2$. Further decrease of the Ce/U content results in the loss of the
long-range character of the magnetic order at finite temperature and in formation of a
characteristic tail in the magnetic phase diagrams. Therefore, no quantum critical phase transition
can be observed in the systems studied. The observed evolution of the magnetic behaviour of the
Ce$_{1-x}$Y$_{x}$NiGe$_{3}$ and U$_{1-x}$Th$_x$NiSi$_2$ is characteristic of the diluted magnetic
alloys. Interestingly, the observed magnetic behaviour exhibits also some similarities to many
solid solutions with fully occupied $f$-electron sublattices \cite{sereni}. Further experiments are
needed to clarify possible relationships between the latter systems and the two solid solutions
studied.

\ack APP thanks D. Kaczorowski for helpful conversations. This work was supported by the Polish
Ministry of Science and Higher Education within research grant no.~N~N202~102338.

\section*{References}
\bibliographystyle{elsarticle-num}

\begin{thebibliography}{10}
\expandafter\ifx\csname url\endcsname\relax
  \def\url#1{\texttt{#1}}\fi
\expandafter\ifx\csname urlprefix\endcsname\relax\def\urlprefix{URL }\fi \expandafter\ifx\csname
href\endcsname\relax
  \def\href#1#2{#2} \def\path#1{#1}\fi

\bibitem{stewart1} Stewart G R 2001 \emph{Rev. Mod. Phys.} \textbf{73} 797

\bibitem{stewart2} Stewart G R 2006 \emph{Rev. Mod. Phys.} \textbf{78} 743

\bibitem{loehneysen} von~L\"{o}hneysen H, Rosch A, Vojta M, W\"{o}lfle P 2007 \emph{Rev. Mod.
    Phys.} \textbf{79} 1015

\bibitem{miranda} Miranda E, Dobrosavljevi\'{c} V 2005 \emph{Rep. Prog. Phys.} \textbf{68} 2337

\bibitem{pikul7} Pikul A P, Caroca-Canales N, Deppe M, Gegenwart P, Sereni J G, Geibel C, Steglich
    F 2006 \emph{J. Phys.: Condens. Matter} \textbf{18} L535

\bibitem{westerkamp} Westerkamp T, Deppe M, K\"{u}chler R, Brando M, Geibel C, Gegenwart P, Pikul
    A P, Steglich F 2009 \emph{Phys. Rev. Lett.} \textbf{102} 206404

\bibitem{pikul8} Pikul A P, Geibel C, Oeschler N, Macovei M E, Caroca-Canales N, Steglich F
    2010 \emph{Phys. Status Solidi B} \textbf{247} 691

\bibitem{pikul9} Pikul A P, Stockert U, Steppke A, Cichorek T, Hartmann S, Caroca-Canales N,
    Oeschler N, Brando M, Geibel C, Steglich F 2012 \emph{Phys. Rev. Letters} \textbf{108} 066405

\bibitem{pikul10} Pikul A P, Kaczorowski D 2011 \emph{J. Phys.: Condens. Matter} \textbf{23} 456002

\bibitem{pikul11} Pikul A P, Kaczorowski D 2011 \emph{J. Phys. Soc. Jpn.} \textbf{80} SA107

\bibitem{salamakha} Salamakha P, Konyk M, Sologub O, Bodak O 1996 \emph{J. Alloys Compd.}
    \textbf{236} 206

\bibitem{akselrud} Akselrud L G, Yarmolyuk Y P, Gladyshevskii E I 1979 \emph{Visn. L'viv. Derzh.
  Univ., Ser. Khim.} \textbf{21} 18

\bibitem{pikul1} Pikul A, Kaczorowski D, Plackowski T, Czopnik A, Michor H, Bauer E, Hilscher G,
    Rogl P, Grin Y 2003 \emph{Phys. Rev. B} \textbf{67} 224417

\bibitem{nakashima1} Nakashima M, Tabata K, Thamizhavel A, Kobayashi T C, Hedo M, Uwatoko Y,
  Shimizu K, Settai R, {\={O}}nuki Y 2004 \emph{J. Phys.: Condens. Matter} \textbf{16} L255

\bibitem{kotegawa1} Kotegawa H, Takeda K, Miyoshi T, Fukushima S, Hidaka H, Kobayashi T C,
  Akazawa T, Ohishi Y, Nakashima M, Thamizhavel A, Settai R, {\={O}}nuki Y 2006 \emph{J. Phys. Soc. Jpn.} \textbf{75} 044713

\bibitem{harada1} Harada A, Kawasaki S, Mukuda H, Kitaoka Y, Thamizhavel A, Okuda Y, Settai R,
    {\={O}}nuki Y, Itoh K M, Haller E E, Harima H 2007 \emph{J. Magn. Magn. Mater.} \textbf{310} 614

\bibitem{harada2} Harada A, Mukuda H, Kitaoka Y, Thamizhavel A, Okuda Y, Settai R, {\={O}}nuki Y,
    Itoh K M, Haller E E, Harima H 2008 \emph{Physica B} \textbf{403} 1020

\bibitem{harada3} Harada A, Mukuda H, Kitaoka Y, Thamizhavel A, Okuda Y, Settai R,
  {\={O}}nuki Y, Itoh K M, Haller E E, Harima H 2008 \emph{J. Phys. Soc. Jpn.} \textbf{77} 103710

\bibitem{pikul6} Pikul A P, Gnida D 2011 \emph{Solid State Commun.} \textbf{151} 778

\bibitem{kaczorowski2} Kaczorowski D 1996 \emph{Solid State Commun.} \textbf{99} 949

\bibitem{pikul12} Pikul A P, Kaczorowski D 2012 \emph{Phys. Rev. B} \textbf{85} 045113

\bibitem{sidorov} Sidorov V A, Tobash P H, Wang C, Scott B L, Park T, Bauer E D, Ronning F,
    Thompson J D, Fisk Z 2011 \emph{J. Phys.: Conf. Series} \textbf{273} 012014

\bibitem{pikul2} Pikul A, Kaczorowski D, Michor H, Rogl P, Bauer E, Hilscher G, Grin Y 2003
    \emph{J. Phys.: Condens. Matter} \textbf{15} 8837

\bibitem{tateiwa} Tateiwa N, Haga Y, Matsuda T D, Ikeda S, Nakashima M, Thamizhavel A, Settai R,
    {\={O}}nuki Y 2006 \emph{J. Phys. Soc. Jpn.} \textbf{75 Suppl.} 174

\bibitem{pikul5} Pikul A P, Kaczorowski D, Michor H, Rogl P, Czopnik A, Grin Y, Bauer E,
    Hilscher G 2003 \emph{Acta Phys. Pol. B} \textbf{34} 1235

\bibitem{mydosh} Mydosh J A, Spin glasses: an experimental introduction, Taylor \& Francis,
  London--Washington, 1993

\bibitem{marcano} Marcano N, Espeso J I, G\'{o}mez Sal J C, Fern\'{a}ndez J R,
  Herrero-Albillos J, Bartolom\'{e} F 2005 \emph{Phys. Rev. B} \textbf{71} 134401

\bibitem{sereni} Sereni J G 1998 \emph{J. Phys. Soc. Jpn.} \textbf{67} 1767

\end{thebibliography}


\end{document}